\documentclass[twocolumn,superscriptaddress,notitlepage]{revtex4-2}

\usepackage{amsmath,amssymb,setspace}
\usepackage{graphicx}
\usepackage[normalem]{ulem}
\usepackage{newtxtext}
\usepackage{newtxmath}
\usepackage[x11names]{xcolor}
\usepackage[unicode,colorlinks,citecolor=Blue3,linkcolor=Blue3,bookmarks=true]{hyperref}

\renewcommand{\Re}{\mathop{\rm Re}}

\newcommand{\fulld}[2]{\dfrac{d#1}{d#2}}
\newcommand{\partd}[2]{\dfrac{\partial#1}{\partial#2}}
\newcommand{\intinfty}{\displaystyle\int_{-\infty}^{\infty}\!}
\newcommand{\bra}[1]{\langle#1|}
\newcommand{\ket}[1]{|#1\rangle}
\newcommand{\bracket}[2]{\langle#1|#2\rangle}
\newcommand{\mean}[1]{\langle#1\rangle}
\newcommand{\hc}{\mathrm{h.c.}}
\newcommand{\hamilt}{\hat{\mathcal{H}}}
\newcommand{\evol}{\hat{\mathcal{U}}}

\begin{document}


\title{QND measurements of photon number in monolithic microcavities}

\author{S.~N.~Balybin}

\email{sn.balybin@physics.msu.ru}

\affiliation{Russian Quantum Center, Skolkovo IC, Bolshoy Bulvar 30, bld. 1, Moscow, 121205, Russia}

\affiliation{M.V. Lomonosov Moscow State University, Faculty of Physics, Leninskie Gory, 1 bld.2  119991, Russia}

\author{A.~B.~Matsko}

\affiliation{Jet Propulsion Laboratory, California Institute of Technology, 4800 Oak Grove Drive, Pasadena, California 91109-8099, USA}

\author{F.~Ya.~Khalili}

\affiliation{Russian Quantum Center, Skolkovo IC, Bolshoy Bulvar 30, bld. 1, Moscow, 121205, Russia}

\affiliation{NTI Center for Quantum Communications, National University of Science and Technology MISiS, Leninsky prospekt 4, Moscow 119049, Russia}

\author{D.~V.~ Strekalov}
\author{V.~S.~Iltchenko}

\affiliation{Jet Propulsion Laboratory, California Institute of Technology, 4800 Oak Grove Drive, Pasadena, California 91109-8099, USA}

\author{A.~A.~Savchenkov}

\affiliation{Rockley Photonics, 234 E Colorado Blvd Ste 600, Pasadena, CA 91101, USA}

\author{N.~M.~Lebedev}

\affiliation{Russian Quantum Center, Skolkovo IC, Bolshoy Bulvar 30, bld. 1, Moscow, 121205, Russia}

\affiliation{M.V. Lomonosov Moscow State University, Faculty of Physics, Leninskie Gory, 1 bld.2  119991, Russia}

\author{I.~A.~Bilenko}

\affiliation{Russian Quantum Center, Skolkovo IC, Bolshoy Bulvar 30, bld. 1, Moscow, 121205, Russia}

\affiliation{M.V. Lomonosov Moscow State University, Faculty of Physics, Leninskie Gory, 1 bld.2  119991, Russia}

\date{\today}

\begin{abstract}
We revisit the idea of quantum nondemolition measurement (QND) of optical quanta via a resonantly enhanced Kerr nonlinearity taking into account quantum back action and show that the monolithic microcavities enable QND measurement of number of quanta in a weak signal field using a spatially overlapping classical probe field. Due to the cross-phase modulation effect, the phase of the probe field acquires information about the signal number of quanta without altering it. We find the exact solution to the Heisenberg equations of motion of this system and calculate the measurement error, accounting for the optical losses in the measurement path. We identify a realistic approximation to obtain the explicit form of the final conditional quantum state of the signal field, accounting for the undesirable self-phase modulation effect and designing the optimal homodyne measurement of the probe beam to evade this effect. We show that the best modern monolithic microcavities allow achieving the measurement imprecision several times better than the standard quantum limit.
\end{abstract}

\maketitle

\section{Introduction}

According to the von Neumann's reduction postulate \cite{Neumann_e}, any ideal (that is precise and free from technical imperfections) quantum measurement leaves the object in the eigen state of the measured observable corresponding the eigen value obtained as the measured result. If the object is already in such an eigen state before the measurement, then this state remains unchanged after the measurement. The majority of the real world measurement devices does not obey this rule perturbing the measured observable.

Let us consider a problem of a non-disturbing detection of a number of photons in a mode of a lossless optical resonator, as an example. A standard tool for the measurement of the number of photons, a photocounter, absorbs all the counted photons, leaving the optical cavity field in the ground state. An output of a phase-preserving linear amplifier \cite{Heffner_ProcIRE_50_1604_1962}, that also can be utilized for the photon counting, depends  on the two non-commuting quadratures of an input mode and, due to the Heisenberg uncertainty relation, disturbs them. It can not measure the number of quanta $n$ in this mode with the precision better than the standard quantum limit (SQL)
\begin{equation}\label{SQL}
  \Delta n_{\rm SQL} = \sqrt{\bar{n}} \,,
\end{equation}
where $\bar{n}$ is the mean number of quanta. Due to the same reasons, it also perturbs $n$  by at least the same value.

A sufficient condition for implementation the ideal von Neumann's measurement was explicitly formulated by David Bohm \cite{Bohm1951}. He showed that the eigen states of the measured observable $\hat{q}$ are not affected by the measurement if this observable commutes with the Hamiltonian $\hat{H}$ of the combined system:
\begin{equation}\label{QND_cond_1}
  [\hat{q},\hat{H}] = 0 \,,
\end{equation}
where
\begin{equation}
  \hat{H} = \hat{H}_S + \hat{H}_A + \hat{H}_I \,,
\end{equation}
$\hat{H}_S$, $\hat{H}_A$ are, respectively, the Hamiltonians of the object and the meter, and $\hat{H}_I$ is the interaction Hamiltonian (it was assumed in \cite{Bohm1951}, that during the measurement, $\hat{H}_{S,A}\to0$, but this assumption does not affect the main conclusion). The term {\it Quantum Non-Demolition} (QND) measurements was proposed for this class of quantum measurements in the late 70s \cite{74a1eBrVo, 80a1BrThVo} and has become generally accepted since then.

The number of quanta in an electromagnetic mode of a linear cavity is an integral of motion, commuting with the Hamiltonian of the mode, $[\hat{n},\hat{H}_S]=0$ (the {\it QND observable}). In this case, the condition \eqref{QND_cond_1} can be simplified to the commutativity with the interaction Hamiltonian:
\begin{equation}\label{QND_cond_2}
  [\hat{n},\hat{H}_I] = 0 \,.
\end{equation}
In other words, the coupling of the mode with the meter has to be non-linear in the mode generalized coordinate represented by the field strength.

Following the initial semi-gedanken proposals \cite{78a1eBrKhVo, Unruh_PRD_18_1764_1978, 80a1eBrKh}, a realistic scheme of QND measurement of electromagnetic photons number was proposed \cite{81a1eBrVy, Milburn_PRA_28_2065_1983}. It  uses two spatially overlapping optical modes, the signal (object) and the meter (probe) ones, interacting by means of the cubic optical nonlinearity $\chi^{(3)}$ arising due to the Kerr effect. During the interaction, the phase of the probe mode is shifted by the value proportional to the photon number in the signal mode. This effect is called {\it cross phase modulation} (XPM). The photon number of the signal mode is preserved during the interaction in the ideal lossless case. The phase of this mode is perturbed by the number of quanta uncertainty of the probe mode due to the same XPM mechanism. This perturbation ensures the fulfillment of the Heisenberg uncertainty principle meaning that reduction of the uncertainty of a signal observable should lead to increase of the uncertainty of an observable conjugated to the signal.

Later on, a significant amount of experimental work based on this idea was done, starting form the pioneering work \cite{Levenson_PRL_57_2473_1986}, see the reviews \cite{Roch_APB_55_291_1992, 96a1BrKh, Grangier_Nature_396_537_1998} for more detail. The sensitivity exceeding the SQL was demonstrated in these experiments, but the single-photon accuracy limit was not reached.

In parallel, another class of the QND schemes, which uses single atoms \cite{Raimond_RMP_73_565_2001, Reiserer_RMP_87_1379_2015, Niemietz_Nature_591_570_2021} or superconductive nonlinear circuits (artificial atoms) \cite{XiuGu_PhysRep_718_1_2017} as nonlinear elements, was actively developed during the last decades. These, in essence lumped devices, are capable of providing resonant cubic nonlinearity many orders of magnitude larger than the electronic nonlinearity of the transparent dielectrics, which let to successful measurements of single quanta.

An important disadvantage of the atom-based QND measurements is their complexity. It is desirable to perform the measurements on a chip, without involvement of bulky equipment needed for the atomic systems. The superconductive circuits can be, and usually are, implemented on-chip, but they operate in a microwave band and require cryogenic cooling. In addition, both these classes, while sensitive to single or a few quanta, do not scale well to bright (multi-quanta) states.

The main problem with the pure optical implementations of QND is that the high optical nonlinearity is typically associated with the high absorption. The promising way to overcome this problem is the usage of whispering-gallery-mode optical (WGM) microresonators \cite{89a1BrGoIl, Strekalov_JOptics_18_123002_2016}, which combine very high quality ($Q-$) factors, exceeding  $10^{11}$ in crystalline microresonators \cite{Savchenkov_OE_15_6768_2007} and $10^9$ in on-chip ones \cite{Kippenberg_PRL_93_083904_2004, Wu_OL_45_5129_2020}, with high concentration of the optical energy in the small volume of the optical modes.

Another major problem, specific to the $\chi^{(3)}$ nonlinearity, is associated with the {\it self phase modulation} (SPM) effect resulting in perturbation of the phases of both the probe and signal modes by the energy uncertainties of the corresponding modes \cite{Imoto_PRA_32_2287_1985}. In it not so crucial for the signal mode, because, while distorting (squeezing) its final quantum state, it does not affect the number of quanta in this mode. At the same time, it introduces an additional uncertainty into the phase of probe mode, proportional to the number of quanta uncertainty in this mode, thus limiting the measurement precision (see details in Sec.\,\ref{sec:simple}). This is so called quantum back action effect.

Two straightforward methods of cancellation of this effect were proposed in Ref.\,\cite{Imoto_PRA_32_2287_1985}: either using a resonant $\chi^{(3)}$ medium or passing the probe beam  through a negative $\chi^{(3)}$ medium before the detection. More recently,  implementations of optical QND measurements using rubidium atoms in a magneto-optical trap were studied experimentally \cite{Roch_PRL_78_634_1997}. It was also noticed that semiconductor quantum dots can provide the negative nonlinearity of proper magnitude to compensate for the SPM in experiments with quantum solitons \cite{Matsko_PRL_82_3244_1999}.

Unfortunately, these methods can not be considered as simple ones. A more practical method based on the measurement of the optimal quadrature of the output probe field instead of the phase one, was proposed in Ref.\,\cite{Drummond_PRL_73_2837_1994}. This measurement allows one to eliminate the major linear part of the SPM and can be made using the ordinary homodyne detectors.

A QND method involving an optimal detuning of both the input field frequencies from the corresponding resonance frequencies of the optical modes was proposed in Ref.\,\cite{Xiao_OE_16_21462_2008}. However, as it follows from that paper, the detuning optimization cancels not only the SPM in the probe beam, but also the perturbation of the signal mode phase by the probe beam number of quanta uncertainty due to the XPM effect, thus violating the Heisenberg uncertainty relation \eqref{Heisenberg}.

In Ref.\,\cite{Tyc_NJP_10_032041_2008}, a theoretical analysis of the optical modes coupled by means of the XPM effect was done, showing that strongly non-Gaussian quantum states of light can be prepared using such systems. However, that work was aimed only at the preparation of the signal mode quantum state for the particular case of the initial coherent state (the term ``QND'' was not used in the paper at all). Also, the SPM effect was not taken into account, and only linearized treatment was performed.

In this paper we study a theoretical feasibility of an ideal measurement of photon quanta in a nonlinear microcavity. Following previous developments in the field, we
(i) provide the consistent quantum analysis of the two-mode QND measurement scheme based on the $\chi^{(3)}$-nonlinearity, accounting for the SPM effect, and (ii) evaluate the prospects of experimental implementation of this scheme using the recent achievements in fabrication of the high-$Q$ WGM resonators. As the result of our analysis it is possible to conclude that the photon number in the probe mode should exceed the photon number in the signal mode if the optical measurement strategy is selected.

This article is organized as follows. In Sec.\,\ref{sec:simple}, we start with the simplified semi-classical treatment of the measurement scheme and find the measurement sensitivity. In Sec.~\ref{sec:Heisenberg}, we derive the exact solution to the corresponding Heisenberg equations of motion and identify an important, from the practical point of view, asymptotic case of this solution. In Sec.~\ref{sec:Schroedinger}, we find the explicit forms of the final quantum state of the signal mode and of the probability distribution for the measurement results for this asymptotic case. In Sec.~\ref{sec:estimates}, we estimate the sensitivity, achievable using the modern WGM microresonators for the QND measurements. Finally, in Sec.~\ref{sec:concl} we summarize the results of this work.

\section{Simplified analysis}\label{sec:simple}

In this section, we use a semi-classical approach considering the classical equations of motion and assuming that the initial values of the involved observables have quantum uncertainties. The validity of the approach will be justified in the next section.

Let us assume that the evolution of the phases $\phi_p$, $\phi_s$ of the probe (p) and the signal (s) waves propagating in a nonlinear media with a cubic (Kerr) nonlinearity in the rotating-wave frame is described by the following equations:
\begin{subequations}\label{simple_phis}
  \begin{gather}
    \phi_p(t) = \phi_p + \Gamma_Sn_p + \Gamma_Xn_s \,, \label{simple_phi_p} \\
    \phi_s(t) = \phi_s + \Gamma_Sn_s + \Gamma_Xn_p \,, \label{simple_phi_s}
  \end{gather}
\end{subequations}
where $n_{p,s}$ are the photon numbers in these modes, which are preserved during the interactions, $\phi_{p,s}$ are the initial values of the phases,
\begin{equation}\label{gammas}
  \Gamma_{S,X} = \gamma_{s,x}\tau \,,
\end{equation}
$\gamma_s$, $\gamma_x$ are the SPM and XPM nonlinearity factors, and $\tau$ is the effective duration of the interaction. The last two terms in Eqs.\,\eqref{simple_phis}, proportional to $\Gamma_X$, describe, respectively, the signal phase shift in the probe mode and the perturbation of the signal mode phase:
\begin{equation}\label{phi_pert_simple}
  \Delta\phi_{s\,{\rm pert}} = \Gamma_X\Delta n_p \,,
\end{equation}
where $\Delta n_p$ is the initial uncertainty of $n_p$.

Suppose that the output phase of the probe mode $\phi_p(t)$ is measured by a phase-sensitive detector. In this case, initial uncertainties of both the phase and the number of quanta of the probe mode contribute to the measurement error. The signal photon number, $n_s$, can be estimated with the uncertainty
\begin{equation}\label{Dn_s_simple}
  \Delta n_{s\,{\rm meas}}
  = \frac{1}{\Gamma_X}\sqrt{(\Delta\phi_p)^2 + \Gamma_S^2(\Delta n_p)^2}
    \ge \frac{\Gamma_S}{\Gamma_X}\Delta n_p \,.
\end{equation}
Since usually $\Gamma_X \sim \Gamma_S$, we find that
\begin{equation}
  \Delta n_{s\,{\rm meas}} \gtrsim \Delta n_p \,.
\end{equation}
For the coherent initial quantum state of the probe mode
\begin{equation}\label{probe_coh}
  \Delta\phi_p = \frac{1}{2\sqrt{\bar{n}_p}} \,, \quad \Delta n_p = \sqrt{\bar{n}_p} \,,
\end{equation}
where $\bar{n}_p$ is the expectation number of probe quanta, Eq.\,\eqref{Dn_s_simple} results in
\begin{equation}\label{Dn_s_simple_coh}
  \Delta n_{s\,{\rm meas}}
  = \frac{1}{\Gamma_X}\sqrt{\frac{1}{4\bar{n}_p} + \Gamma_S^2\bar{n}_p}
   > \frac{\Gamma_S}{\Gamma_X}\sqrt{\bar{n}_p} \,.
\end{equation}
As it follows from this inequality, in order to overcome the SQL (see Eq.~\eqref{SQL}), there should be $\bar{n}_s \gtrsim \bar{n}_p$, which makes high-precision QND measurement of small number of quanta impossible.

Let us consider now a measurement of the linear combination of the probe phase and photon number (see \cite{Imoto_PRA_32_2287_1985})
\begin{equation}
  \phi_p(t) - \Gamma_Sn_p(t) = \phi_p + \Gamma_Xn_s \,.
\end{equation}
In this case, the sensitivity is affected only by initial uncertainty of the probe phase:
\begin{equation}\label{Dn_opt_simple}
  \Delta n_{s\,{\rm meas}} = \frac{\Delta\phi_p}{\Gamma_X} \,,
\end{equation}
which could be arbitrary small, provided sufficiently big non-linearity factor $\Gamma_X$ and the probe photon number.

It follows also form Eqs.\,(\ref{phi_pert_simple}, \ref{Dn_opt_simple}) that
\begin{equation}\label{Heisenberg}
  \Delta n_{s\,{\rm meas}}\Delta\phi_{s\,{\rm pert}} = \Delta\phi_p\Delta n_p
    \ge \frac{1}{2}\,,
\end{equation}
that is, the uncertainty relation for the number of quanta and phase of the probe mode directly translates to the uncertainty relation for $\Delta n_{s\,{\rm meas}}$ and $\Delta\phi_{s\,{\rm pert}}$. In particular, in the case of the coherent quantum state of the probe mode
\eqref{probe_coh},
\begin{subequations}\label{uncertainty}
  \begin{gather}
    \Delta n_{s\,{\rm meas}} = \frac{1}{2\Gamma_X\sqrt{\bar{n}_p}} \,,\label{Dn_opt_s1}\\
    \Delta\phi_{s\,{\rm pert}} = \Gamma_X\sqrt{\bar{n}_p} \,.
  \end{gather}
\end{subequations}

Finally, the necessary condition for a successful sub-SQL measurement can be presented as $\Delta n_{s\ {\rm meas}} < \sqrt{\bar n_s}$. With account for Eq.\,\eqref{Dn_opt_s1}, it corresponds to the following inequality:
\begin{equation}
  2\Gamma_X\sqrt{\bar n_p}\sqrt{\bar n_s} > 1.
\end{equation}

\section{Measurement imprecision}\label{sec:Heisenberg}

Using the rotating-wave approximation, the Hamiltonian of the two modes system, considered in the previous section, can be presented as follows:
\begin{equation}\label{hamilt}
  \hamilt = -\frac{\hbar\gamma_S}{2}\sum_{x=s,p}\hat{n}_x(\hat{n}_x-1)
    - \hbar\gamma_X\hat{n}_p\hat{n}_s \,,
\end{equation}
where $\hbar$ is the reduced Plank constant,
\begin{equation}\label{const_n_sp}
  \hat{n}_{s,p} = \hat{a}_{s,p}^\dag\hat{a}_{s,p}
\end{equation}
and $\hat{a}_{s,p}$, $\hat{a}_{s,p}^\dag$ are the annihilation and creation operators of the signal  and the probe modes (the peculiar form of the first term is the normal-ordered one). It can be seen from this Hamiltonian, that the numbers of quanta in both modes (in the Heisenberg picture) are integrals of motion of the system:
\begin{equation}\label{n_of_t}
  \hat{n}_{s,p}(t) = \hat{n}_{s,p}
\end{equation}

The corresponding Heisenberg equations of motion for the annihilation operators are
\begin{subequations}\label{eqs_a_sp}
  \begin{gather}
    \fulld{\hat{a}_p(t)}{t} = i[\gamma_S\hat{n}_p(t) + \gamma_X\hat{n}_s(t)]\hat{a}_p(t)\,,\\
    \fulld{\hat{a}_s(t)}{t} = i[\gamma_S\hat{n}_s(t) + \gamma_X\hat{n}_p(t)]\hat{a}_s(t) \,.
  \end{gather}
\end{subequations}
Due to the photon number conservation \eqref{n_of_t} the closed form of the solution of the set of equations can be easily found:
\begin{subequations}\label{soln_a_sp}
  \begin{gather}
    \hat{a}_p(t) = e^{i(\Gamma_S\hat{n}_p + \Gamma_X\hat{n}_s)}\hat{a}_p\,,\\
    \hat{a}_s(t) = e^{i(\Gamma_S\hat{n}_s + \Gamma_X\hat{n}_p)}\hat{a}_s \,.
  \end{gather}
\end{subequations}.

Let us consider now the homodyne measurement of the quadrature $\hat{X}_\zeta$ of the probe mode, defined by the homodyne angle $\zeta$:
\begin{equation}
  \hat{X}_\zeta = \frac{1}{\sqrt{2}}\bigl[\hat{a}_p(t)e^{i\zeta} + \hc\bigr]
  = \frac{1}{\sqrt{2}}
      \bigl[e^{i(\Gamma_S\hat{n}_p + \Gamma_X\hat{n}_s + \zeta)}\hat{a}_p + \hc\bigr] ,
\end{equation}
where ``$\hc$'' stands for ``Hermitian conjugate''. The measurement error for the number of quanta in the signal mode can be calculated by standard error propagation formula:
\begin{equation}
  (\Delta n_s)^2 = \frac{(\Delta X_\zeta)^2}{G^2} \,,
\end{equation}
where
\begin{equation}
  G = \partd{\mean{\hat{X}_\zeta}}{n_s}
\end{equation}
is the transfer function,
\begin{equation}
  (\Delta X_\zeta)^2 = \mean{\hat{X}_\zeta^2} - \mean{\hat{X}_\zeta}^2 \,,
\end{equation}
and momenta $\mean{\hat{X}_\zeta}$, $\mean{\hat{X}_\zeta^2}$ are calculated for a given value of $n_S$, that is for the Fock state $\ket{n_s}$ of the signal mode.

If the probe mode state is prepared in a coherent state $\ket{\alpha}_p$, the expectation value of the probe amplitude can be selected to be real
\begin{equation}
  \alpha = \sqrt{\bar{n}_p}.
\end{equation}
The exact value of the measurement error $\Delta n_s$ for this case is calculated in the App.\,\ref{app:Dn_s}, see Eqs.\,\eqref{GD2Xn_gen}.

In the case of weak non-linearity and strong probe field the solution can be simplified. Let us assume that
\begin{equation}\label{approx}
  |\Gamma_S|\to0 \,, \quad \bar{n}_p\to\infty \,, \quad
  \text{but }\Gamma_S\bar{n}_p\text{ remains finite}.
\end{equation}
This approximation is well-satisfied for the realistic whispering gallery mode (WGM) resonators (see Sec.\,\ref{sec:estimates}).
In this case (see App.\,\ref{app:Dn_s})
\begin{subequations}\label{D2X_G_approx}
  \begin{gather}
    (\Delta X)^2 = \frac{1}{2} - \Gamma_S\bar{n}_p\sin2\varphi
      + 2\Gamma_S^2\bar{n}_p^2\sin^2\varphi \,, \label{D2X_approx} \\
    G = -\sqrt{2}\alpha\Gamma_X\sin\varphi \,,
  \end{gather}
\end{subequations}
where
\begin{equation}\label{varphi}
  \varphi = \Gamma_S\bar{n}_p + \Gamma_Xn_s + \zeta \,.
\end{equation}

These equations correspond to the ideal exact measurement of $\hat{X}_\zeta$. The losses in the measurement channel can be taken into account by introducing its unified quantum efficiency $\eta$ (which includes, in particular, the finite quantum efficiency of the homodyne detector) as
\begin{subequations}
  \begin{gather}
    (\Delta X)_\eta^2 = \eta(\Delta X)^2 + \frac{1-\eta}{2} \,, \label{D2X_loss} \\
    G_\eta = \sqrt{\eta}G  \,,
  \end{gather}
\end{subequations}
resulting in an expression for the measurement error for the number of quanta in the signal mode:
\begin{equation}\label{D2n_s_loss}
  (\Delta n_s)^2 = \frac{(\Delta X_\zeta)_\eta^2}{G_\eta^2}
  = \frac{1}{\Gamma_X^2}\biggl[
        \frac{1 + (\cot\varphi - 2\eta\Gamma_S\bar{n}_p)^2}{4\eta\bar{n}_p}
        + (1-\eta)\Gamma_S^2\bar{n}_p
      \biggr] \,.
\end{equation}

Following the reasoning of Sec.\,\ref{sec:simple}, we assume that $\cot\varphi=0$, which corresponds to the maximum of the transfer function as well as to the measurement of the phase quadrature of the output probe beam. Following this path we arrive at
\begin{equation}
  (\Delta n_s)^2 = \frac{1}{\Gamma_X^2}
    \biggl(\frac{1}{4\eta\bar{n}_p} + \Gamma_S^2\bar{n}_p\biggr) .
\end{equation}
In the ideal case of $\eta=1$, this equation reduces to Eqs.\,(\ref{Dn_s_simple_coh}).

The minimum of the measurement error of the optimized detection procedure described by \eqref{D2n_s_loss}
\begin{equation}\label{d2n_s_opt}
  (\Delta n_{s,\rm min})^2 = \frac{1}{\Gamma_X^2}
    \biggl[\frac{1}{4\eta\bar{n}_p} + (1-\eta)\Gamma_S^2\bar{n}_p\biggr] ,
\end{equation}
is achieved at the optimum angle $\varphi$ given by
\begin{equation}\label{varphi_opt}
  \cot\varphi = 2\eta\Gamma_S\bar{n}_p.
\end{equation}
In the ideal case of $\eta=1$, the additional term in $(\Delta n_s)^2$ vanishes, giving Eq.\,\eqref{Dn_opt_s1}.

Our reasoning contains a ``vicious loop'': the value of $\zeta$, defined by Eq.\,\eqref{varphi_opt}, depends on the measured value $n_s$, which is unknown before the measurement. Assuming that $1 \gg |n_s-\bar n_s|/\bar n_s$ and that $\bar n_s$ is known we replace Eq.\,\eqref{varphi_opt} with the following condition
\eqref{varphi_opt}:
\begin{subequations}\label{cot_bar_phi}
  \begin{equation}
    \cot\bar{\varphi} = 2\eta\Gamma_S\bar{n}_p \,,
  \end{equation}
  where
  \begin{equation}\label{bar_varphi}
    \bar{\varphi} = \Gamma_S\bar{n}_p + \Gamma_X\bar{n}_s + \zeta \,.
  \end{equation}
\end{subequations}
It is shown in App.\,\ref{app:eps}, that under reasonable assumptions, the condition \eqref{cot_bar_phi} leads only to a minor correction to Eq.\,\eqref{d2n_s_opt}, see Eq\,.\eqref{d2n_s_mod}.

The optimal number of probe mode quanta, equal to
\begin{equation}\label{n_p_opt}
  \bar{n}_p^{\rm opt} = \frac{1}{2\Gamma_S\sqrt{\eta(1-\eta)}} \,,
\end{equation}
exists and follows form Eq.\,\eqref{d2n_s_opt}. This selection minimizes the measurement error, with the minimum being equal to
\begin{equation}\label{d2n_s_min}
  (\Delta n_{s,min}^{\rm opt})^2 = \frac{\Gamma_S}{\Gamma_X^2}\sqrt{\frac{1-\eta}{\eta}} \,.
\end{equation}

\section{Conditional state of the signal mode}\label{sec:Schroedinger}

Let us consider the wave function of the final quantum state of the joint two-mode systems and assume for simplicity that $\eta=1$. Using the Hamiltonian \eqref{hamilt} we find for the final state
\begin{equation}
  \ket{\Psi} = \evol\ket{\alpha}_p\otimes\ket{\psi}_s \,,
\end{equation}
where
\begin{equation}
  \evol = \exp\frac{\hamilt t}{i\hbar}
  = \exp\biggl[
      \frac{i\Gamma_S}{2}\sum_{x=p,s}\hat{n}_x(\hat{n}_x-1) + i\Gamma_X\hat{n}_p\hat{n}_s
    \biggr]
\end{equation}
is the evolution operator.

Measurement of the probe mode quadrature $\hat{X}_\zeta$ reduces the signal mode into the following quantum state
\begin{equation}
  \ket{\psi(X)} = \frac{\hat{\Omega}(X)\ket{\psi}_s}{\sqrt{W(X)}} \,,
\end{equation}
where $X$ is the measurement result
\begin{equation}
  \hat{\Omega}(X) = {}_p\bra{X,\zeta}\evol\ket{\alpha}_p
  = \sum_{n_s=0}^\infty e^{i\Gamma_Sn(n-1)/2}\Omega(X,n_s)\ket{n_s}_s\,{}_s\bra{n_s}
\end{equation}
is the reduction (Kraus) operator, $\ket{X,\zeta}_p$ is the eigenstate of $\hat{X}_\zeta$ with the eigenvalue $X$, and
\begin{equation}
  W(X) = {}_s\bra{\psi}\hat{\Pi}(X)\ket{\psi}_s
\end{equation}
is the {\it a priory} probability distribution of $X$,
\begin{equation}
  \hat{\Pi}(X) = \hat{\Omega}^\dag(X)\hat{\Omega}(X)
  = \sum_{n_s=0}^\infty|\Omega(X,n_s)|^2\ket{n_s}_s\,{}_s\bra{n_s}
\end{equation}
is the positive operator-valued measure (POVM) \cite{HelstromBook} for this measurement, and
\begin{equation}\label{Omega_Xn}
  \Omega(X,n_s) = {}_p\bra{X,\zeta}
    \exp\biggl[\frac{i\Gamma_S}{2}\hat{n}_p(\hat{n}_p-1) + i\Gamma_Xn_s\hat{n}_p\biggr]
    \ket{\alpha}_p \,.
\end{equation}

The explicit form of the reduction operator  $\hat{\Omega}(X)$ for the asymptotic case of \eqref{approx} is calculated in Appendix \ref{app:Schroedinger}, see Eq.\,\eqref{Omega_approx}. It follows from this result, that the conditional probability distribution of $X$ for a given number of quanta in the signal mode $n$ is equal to
\begin{equation}\label{Omega2}
  |\Omega(X,n)|^2 = \frac{1}{\sqrt{2\pi(\Delta X)^2}}
    \exp\biggl[-\frac{(X - \sqrt{2}\alpha\cos\varphi)^2}{2(\Delta X)^2}\biggr],
\end{equation}
with the values of $(\Delta X)^2$ and $\varphi$  are given by Eqs.\,(\ref{D2X_approx}, \ref{varphi_opt_n}).

The {\it a posteriori} probability distribution for $n_s$ conditioned on the measured value of $X$, can be obtained form Eq.\,\eqref{Omega2} using Bayes' theorem:
\begin{equation}\label{W_apost}
  W_{\rm apost}(n_s|X) = \frac{1}{\mathcal{W}(X)}|\Omega(X,n_s)|^2W_{\rm apr}(n_s) \,,
\end{equation}
where $W_{\rm apr}$ is the {\it a priori} probability distribution and
\begin{equation}
  \mathcal{W}(X) = \sum_{n=0_s}^\infty|\Omega(X,n)|^2W_{\rm apr}(n_s)
\end{equation}
is the normalization factor, equal to the unconditional probability distribution for $X$.

It interesting to note, that the function \eqref{Omega2} is Gaussian in $X$, but non-Gaussian in $n_s$ due to the dependence of $\varphi$ on $n_s$, see Eq.\,\eqref{varphi}. Therefore, the probability distribution \eqref{W_apost} is also non-Gaussian (see also the brief discussion in the end of App.\,\ref{app:eps}).

\begin{figure*}[ht]
\centering
  \includegraphics[width=0.8\linewidth]{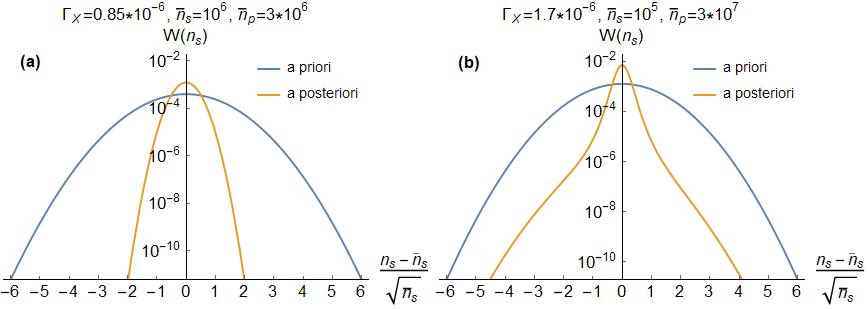}
  \caption{ The {\it a priori} \eqref{W_apr} and {\it a posteriori} \eqref{W_apost} probability distributions for the number of quanta in the signal mode. (a) the parameters, discussed in Sec.\,\ref{sec:estimates}. (b) Slightly increased values of $\Gamma_X$, $\bar{n}_p$, giving non-Gaussian shape of the {\it a posteriori} distribution. In both cases, the initial quantum state of the signal mode is assumed to be a coherent one.}
  \label{fig:n_dist}
\end{figure*}

In Fig.\,\ref{fig:n_dist}, the probaility distribution is plotted as a function of $n_s$. That picture was plotted using the distribution \eqref{W_apost} and assuming the condition \eqref{cot_bar_phi}.  The initial probability distribution, which we assume to be the Poissonian one (corresponding to the coherent initial state of the signal mode):
\begin{equation}\label{W_apr}
  W_{\rm apr}(n) = \frac{e^{-\bar{n}_s}\bar{n}_s^n}{n!}
\end{equation}
is also shown for the comparison. Panel (a) illustrates the result of QND measurement with the parameters close to the realistic experimental values discussed in Sec.\,\ref{sec:estimates}. In the panel (b) we used higher values of $\Gamma_X$, $\bar{n}_p$ which give non-Gaussian shape of the {\it a posteriori} distribution.

\section{Discussion}\label{sec:estimates}

Let us evaluate the efficiency and requirements of the QND measurements performed with high-Q WGM resonators. The factors $\Gamma_X$, $\Gamma_S$ [see Eq.\,\eqref{gammas}] for the XPM and SPM effects based on the electronic nonlinearity of the material can be estimated as
\begin{equation}
 	\Gamma_X = 2\Gamma_S
   = 2Q_{\rm load}\frac{n_2}{n_0} \frac{\hbar \omega_0 c}{{V_{\rm eff}}}\,.
\end{equation}
where $c$ is the speed of light, $\omega_0$ is the optical frequency, $n_0$ is the refractive index of the material, $n_2$ is the cubic nonlinearity coefficient, $V_{\rm eff}$ is the effective volume of the mode, and $Q_{\rm load}=\omega_0\tau$ is the loaded quality factor. Note that one of the factors, which constitute the unified quantum efficiency $\eta$, is equal to
\begin{equation}
  \eta_{\rm load} = 1 - \frac{Q_{\rm load}}{Q_{\rm intr}} \,.
\end{equation}
where $Q_{\rm intr}$ is the intrinsic quality factor. Therefore, in order to overcome the SQL by a significant margin, $Q_{\rm load}$ should be smaller that $Q_{\rm intr}$ by 1-2 orders of magnitude.

We select ${\rm CaF}_2$ as the resonator host material in which the highest quality factor $Q_{\rm intr}=3\times10^{11}$ was achieved so far \cite{Savchenkov_OE_15_6768_2007}. We assume the that the (vacuum) wavelengths are close to $\lambda = 2\pi c/\omega_0 = 1.55\,\mu{\rm m}$ for both the signal and probe modes. At this wavelength, ${\rm CaF}_2$ is characterized by the refractive index $n_0=1.44$ and the nonlinearity factor $n_2 = 3.2 \times 10^{-20}\,{\rm m^2/W}$. We assume also that the resonator has $100~\mu{\rm m}$ in diameter. The circumference of the resonator is shaped in a sharp edge resulting in $2\,\mu{\rm m}\times3\,\mu{\rm m}$ mode cross-section and, correspondingly, $V_{\rm eff}\simeq 2 \times 10^{-15}\,{\rm m^3}$ mode volume.

For these parameters, the factors $\Gamma_X$ and $\Gamma_S$ can be estimated as follows:
\begin{equation}
 	\Gamma_X = 2\Gamma_S \approx 0.85\times10^{-6}\times\frac{Q_{\rm load}}{10^9} \,.
\end{equation}
For a reasonably optimistic value of $\eta=0.9$, these parameters translate to the following values of the optimal number of the probe quanta and the corresponding measurement error, see Eqs.\,(\ref{n_p_opt}, \ref{d2n_s_min}):
\begin{gather}
  \bar{n}_p^{\rm opt} \approx 4\times10^6\times\frac{10^9}{Q_{\rm load}} \,,\label{n_p_est}\\
  (\Delta n_{s,\rm min}^{\rm opt})^2 \approx 2\times10^5\times\frac{10^9}{Q_{\rm load}} \,.
\end{gather}
The pump power which is necessary to excite the intracavity number of quanta \eqref{n_p_est}, can be estimated as follows:
\begin{equation}
	P_p = \frac{\hbar\omega_0^2\bar{n}_p}{2Q_{\rm load}}
  \approx 0.3\,\mu{\rm W}\times\biggl(\frac{10^9}{Q_{\rm load}}\biggr)^2 \,.
\end{equation}

This estimation shows that we can create a nonclassical state of the signal field using several times smaller number of the signal photons (for example, $\bar{n}_s\approx10^6$), than the number of probe photons.

\begin{figure*}[t]
\centering
  \includegraphics[width=0.8\linewidth]{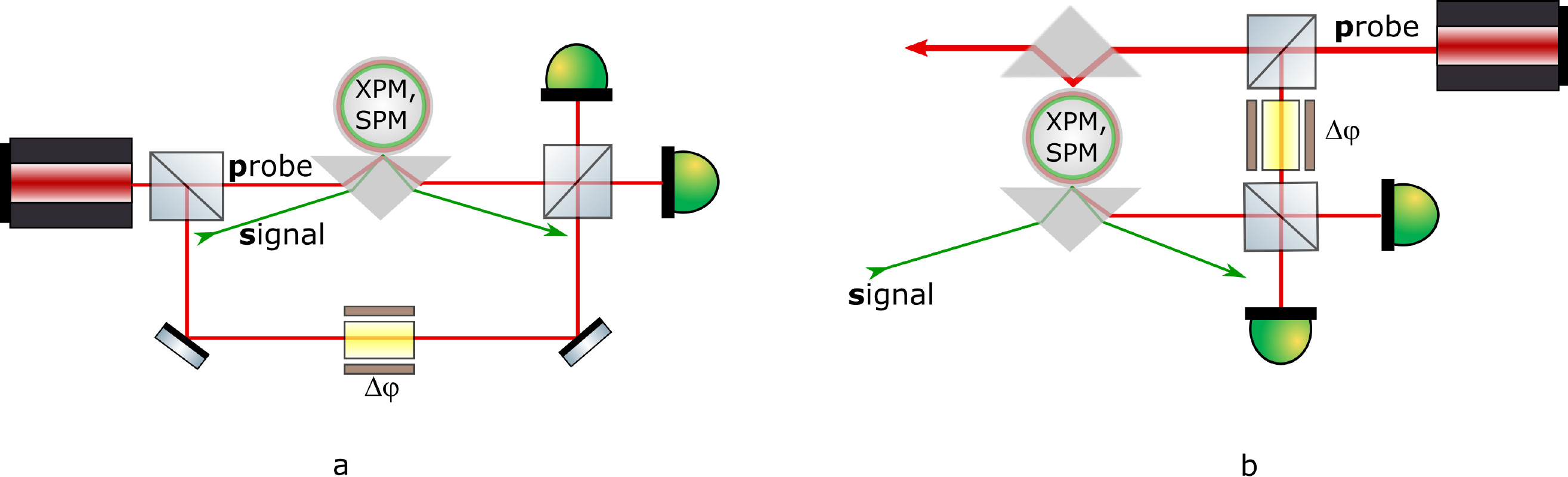}
  \caption{Possible experimental implementations of QND detection using WGM monolithic microresonators equipped with, (a), a single evanescent field coupler and, (b), two  couplers. }
  \label{fig:exp_setup}
\end{figure*}

Possible conceptual implementations of the proposed measurement are illustrated by Fig.\,(\ref{fig:exp_setup}a). We assume that wavelengths of the probe and signal waves are dissimilar enough to allow their separation outside of the resonator. Probe wave (laser output in a coherent state) is injected to the resonator by means of a prism coupler. Part of the laser output is optimally phase shifted and used as a reference. Signal wave is coupled to the resonator using the same prism. A quadrature component of the probe emitted from the resonator is measured using a balanced homodyne detector. In order to preserve quantum states of the signal and the probe the resonator modes are overcoupled.  Alternatively, classical probe wave can be injected through an additional, weakly coupled channel (\ref{fig:exp_setup}b). This later approach is preferable because the resonator filters out the unwanted frequency as well as spatial components from the probe.

\section{Conclusion}\label{sec:concl}

Sensing an optical signal by means of a probe wave via cross-phase modulation in a non-linear media paves a way to a non-absorbing measurement of the signal quanta number with the precision beating, under certain conditions, the standard quantum limit. In this paper we show that the state-of-the-art measurements based on the whispering gallery microresonator with Kerr non-linearity and record quality factor may achieve this goal.

We obtained an exact solution to the two mode nonlinear interaction problem using the Heisenberg equations of motion. Our analysis indicates that the self-phase modulation does not limit the measurement accuracy. We found that using a WGM resonator having $100\,\mu{\rm m}$ in diameter, made of pure calcium fluorite, and interrogated with microwatt level probe wave in the $1.5 \,\mu m$ telecom band allows us to detect $n_s\simeq 10^6$ number of quanta with imprecision several times smaller than $\sqrt{n_s}$. We found the explicit form of the conditional final state of the signal mode and shown that it could have a non-Gaussian shape.

These results can be considered as important steps towards the experimental realization of a QND measurement and non-classical state preparation in optical microresonators.  It is worth to note that the recent advances in the integrated technology enable on-chip microresonators with quality factor above $10^8$  \cite{Jin_NPhys_15_346_2021, Wu_OL_45_5129_2020}, so mass production of devices suitable for quantum measurements is not a distant future.

\acknowledgments

The work of S.N.B., F.Y.K. and I.A.B. was supported by the Russian Science Foundation (project 20-12-00344). The work of S.N.B. was also supported by the Basis Foundation (project no. 19-2-6-68-1). The research performed by A.B.M., D.V.S., and V.S.I was carried out at the Jet Propulsion Laboratory, California Institute of Technology, under a contract with the National Aeronautics and Space Administration (80NM0018D0004).

\appendix

\section{Hamiltonian for a cavity with cubic nonlinearity} \label{app:Ham}

A generalized frequency independent third order nonlinear interaction of modes of a nonlinear resonator can be described by a model Hamiltonian \cite{Strekalov_JOptics_18_123002_2016}
\begin{equation}
  \hamilt =\frac{\hbar \gamma_S}{2}  \sum \limits_{i,j,k,l} \hat a_i^\dag \hat a_j^\dag \hat a_k \hat a_l,
\end{equation}
where we introduce a coupling parameter $\gamma_S$ obtained under the assumption of complete spatial overlap of the resonator modes.  For the case of the two interacting resonator modes the Hamiltonian can be simplified  (see Eq.~\ref{hamilt})
\begin{equation} \label{vsimp}
 \hamilt = -\frac{\hbar \gamma_S}{2}  (\hat a_p^\dag (t)+\hat a_s^\dag (t))^2  (\hat a_p (t)+\hat a_s(t))^2 ,
\end{equation}

Equations describing the evolution of the field operators of the resonator modes are generated using the Hamiltonian (\ref{vsimp})
\begin{equation} \label{set}
\dot{\hat a}_j=-i\omega_j\hat a_j+ \frac{i}{\hbar} [\hamilt,\hat a_j],
\end{equation}
where $j=p, s$ and $\omega_j$ are the frequencies of the modes. The set (\ref{set}) can be written in the explicit rotation wave approximation form as
\begin{eqnarray}
\dot{\hat a}_p=-i\omega_p\hat a_p+ i\gamma_S\hat a_p^\dag \hat a_p^2+2i\gamma_S\hat a_s^\dag \hat a_s \hat a_p, \\
\dot{\hat a}_s=-i\omega_s\hat a_s+ i\gamma_S\hat a_s^\dag \hat a_s^2+2i\gamma_S\hat a_p^\dag \hat a_p \hat a_s,
\end{eqnarray}
which explicitly shows (compare with Eqs.~(\ref{eqs_a_sp}) written for slow amplitudes $\hat a_j(t)\exp(i\omega_j t)$) that $\gamma_X$, introduced in the paper, is twice bigger than $\gamma_S$. It worth noting that this relationship can be violated in a resonant nonlinear medium.

\section{Calculation of the measurement error}\label{app:Dn_s}

The straightforward calculation gives that for the initial state of the two-mode system, equal to $\ket{\alpha}_p\otimes\ket{n}_s$, the first two momenta of $X_\zeta$ are equal to
\begin{subequations}
  \begin{gather}
    \mean{\hat{X}_\zeta} = \sqrt{2}\alpha\Re\bigl[
        e^{i(\Gamma_Xn_s + \zeta)}
        {}_p\bra{\alpha}e^{i\Gamma_S\hat{n}_p}\ket{\alpha}_p
      \bigr] , \\
    \mean{\hat{X}_\zeta^2} = \alpha^2\Re\bigl[
        e^{2i(\Gamma_Xn_s + \zeta + \Gamma_S/2)}
          {}_p\bra{\alpha}e^{2i\Gamma_S\hat{n}_p}\ket{\alpha}_p
      \bigr]
      + \alpha^2 + \frac{1}{2} \,.
  \end{gather}
\end{subequations}
Taking into that for any factor $\lambda$,
\begin{equation}\label{lemma_1}
  \bra{\alpha}e^{i\lambda\hat{n}}\ket{\alpha} = \exp[|\alpha|^2(e^\lambda-1)] \,,
\end{equation}
see \cite{Louisell_e}, we obtain:
\begin{subequations}
  \begin{gather}
    \mean{\hat{X}_\zeta} = \sqrt{2}\alpha E_1\cos\varphi \,, \\
    \mean{\hat{X}_\zeta^2} = \alpha^2E_2\cos(2\varphi + \Delta) + \alpha^2 + \frac{1}{2} \,,
  \end{gather}
\end{subequations}
where
\begin{subequations}
  \begin{gather}
    E_1 = \exp[\alpha^2(\cos\Gamma_S-1)] \,, \\
    E_2 = \exp[\alpha^2(\cos2\Gamma_S-1)] \,, \\
    \varphi = \alpha^2\sin\Gamma_S + \Gamma_Xn_s + \zeta \,, \\
    \Delta = \alpha^2(\sin2\Gamma_S - 2\sin\Gamma_S) + \Gamma_S \,.
  \end{gather}
\end{subequations}
Therefore,
\begin{subequations}\label{GD2Xn_gen}
  \begin{gather}
    G = -\sqrt{2}\alpha\Gamma_XE_1\sin\varphi \,, \\
    (\Delta X_\zeta)^2 = \mean{\hat{X}_\zeta^2} - \mean{\hat{X}_\zeta}^2
      = A + B\cos2\varphi - C\sin2\varphi\,,
  \end{gather}
  \begin{multline}\label{D2n_s_gen}
    (\Delta n_s)^2 = \frac{(\Delta X_\zeta)^2}{G^2} \\
      = \frac{1}{2\alpha^2\Gamma_X^2E_1^2}[(A+B)\cot^2\varphi - 2C\cot\varphi + A-B] \,,
  \end{multline}
\end{subequations}
where
\begin{subequations}
  \begin{gather}
    A = \frac{1}{2} + \alpha^2(1 - E_1^2) \,, \\
    B = \alpha^2(E_2\cos\Delta - E_1^2) \,, \\
    C = \alpha^2E_2\sin\Delta \,.
  \end{gather}
\end{subequations}

In the asymptotic case \eqref{approx},
\begin{equation}\label{ABC_app}
  A \to \frac{1}{2} + \alpha^4\Gamma_S^2 \,, \quad
  B \to - \alpha^4\Gamma_S^2 \,, \quad
  C \to \alpha^2\Gamma_S \,,
\end{equation}
which gives Eqs.\,(\ref{D2X_G_approx}, \ref{varphi}).

\section{Accounting for the initial uncertainty of $n_s$}\label{app:eps}

It follows from Eqs.\,(\ref{varphi}, \ref{bar_varphi}), that
\begin{equation}\label{varphi_opt_n}
  \varphi = \bar{\varphi} + \Gamma_X(n_s-\bar{n}_s) \,.
\end{equation}
Let us consider an important, from the practical point of view, case of the initial state of the signal mode which is close to the coherent one:
\begin{equation}\label{near_coher}
  \mean{(n_s-\bar{n}_s)^2} \sim n_s \,.
\end{equation}
Let us assume that
\begin{equation}\label{assumpt2}
  \Gamma_X\sim\Gamma_S \,, \quad \bar{n}_s\lesssim\bar{n}_p  \,.
\end{equation}
Taking into account the approximation \eqref{approx}, one can obtain
\begin{equation}\label{small_dns}
  \Gamma_X|n_s-\bar{n}_s|\ll1 \,.
\end{equation}
Therefore,
\begin{equation}\label{cot_varphi_opt}
  \cot\varphi
  \approx \cot\bar{\varphi} - \frac{\Gamma_X(n_s-\bar{n}_s)}{\sin^2\bar{\varphi}}
  = 2\eta\Gamma_S\bar{n}_p - \epsilon \,,
\end{equation}
where
\begin{equation}\label{eps}
  \epsilon = (1 + 4\eta^2\Gamma_S^2\bar{n}_p^2)\Gamma_X(n_s-\bar{n}_s) \,.
\end{equation}
Substitution of \eqref{cot_varphi_opt} into Eq.\,\eqref{D2n_s_loss} leads to
\begin{equation}\label{d2n_s_mod}
  (\Delta n_s)^2 = \frac{1}{\Gamma_X^2}
    \biggl[\frac{1 + \epsilon^2}{4\eta\bar{n}_p} + (1-\eta)\Gamma_S^2\bar{n}_p\biggr] \,.
\end{equation}
If the number of the probe mode quanta is equal to the optimal one \eqref{n_p_opt}, then  Eq.\,\eqref{eps} reduces to
\begin{equation}
  \epsilon = \frac{\Gamma_X(n_s-\bar{n}_s)}{1-\eta} \,.
\end{equation}

For the values of $\Gamma_X$ and $\eta$, introduced in Sec.\,\ref{sec:estimates}, this corresponds to
\begin{equation}
  \epsilon \approx 0.01\times\frac{n_s-\bar{n}_s}{10^3} \,.
\end{equation}
Therefore, if the assumptions (\ref{approx}, \ref{assumpt2}) are fulfilled and $(n_s-\bar{n}_s)^2\lesssim\mean{(n_s-\bar{n}_s)^2} \sim \bar{n}_s$, then $\epsilon^2\ll1$, which reduces Eq.\,\eqref{d2n_s_mod} to \eqref{d2n_s_opt}. However, if the measured value $n_s$ deviates strongly form the mean value $\bar{n}_s$, then the factor $\epsilon$ becomes significant, introducing dependence on $n_s$ into $\Delta n_s$ and, therefore, making the {\it a posteriori} distribution \eqref{W_apost} non-Gaussian.

\section{Calculation of statistics of the measurement results}\label{app:Schroedinger}

Absorbing the regular phase shifts into $\ket{X,\zeta}$, the kernel \eqref{Omega_Xn} can be presented as follows:
\begin{equation}
  \Omega(X,n)
  = {}_p\bra{X,\varphi}
      \exp\biggl\{\frac{i\Gamma_S}{2}[\hat{n}_p(\hat{n}_p-1) - 2\alpha^2\hat{n}_p]\biggr\}
      \ket{\alpha} \,,
\end{equation}
where
\begin{equation}
  \ket{X,\varphi}_p = e^{-i(\Gamma_S\alpha^2 + \Gamma_Xn)\hat{n}_p}\ket{X,\zeta}_p
\end{equation}
and $\varphi$ is given by Eq.\,\eqref{varphi}. Then rewrite it as follows:
\begin{multline}
  \Omega(X,n) = {}_p\bra{X,\varphi}\hat{\mathcal{D}}(\alpha) \\
    \times\exp\biggl\{\frac{i\Gamma_S}{2}[
      (\hat{a}_p^\dagger{}+\alpha)^2(\hat{a}_p+\alpha)^2
      - 2\alpha^2(\hat{a}_p^\dag+\alpha)(\hat{a}_p+\alpha)
    ]\biggr\} \\
    \times\hat{\mathcal{D}}^\dag(\alpha)\ket{\alpha}_p \,,
\end{multline}
where $\hat{\mathcal{D}}$ is the unitary displacement operator defined by
\begin{equation}
  \hat{\mathcal{D}}^\dag(\alpha)\hat{a}_p\hat{\mathcal{D}}(\alpha) = \hat{a}_p + \alpha \,.
\end{equation}

Assume the approximation \eqref{approx}. In this case, omitting the non-physical factor $e^{-i\Gamma_S\alpha^4/2}$, we obtain:
\begin{multline}
  \Omega(X_\zeta,n) \approx {}_p\bra{X,\varphi}\mathcal{D}(\alpha)
      \exp\biggl[\frac{i\Gamma_S\alpha^2}{2}(\hat{a}_p + \hat{a}_p^\dagger)^2\biggr]
    \mathcal{D}^\dag(\alpha)\ket{\alpha}_p \\
  = {}_p\bra{X,\varphi}e^{i\Gamma_S\alpha^2(\hat{X}-\alpha\sqrt{2})^2}\ket{\alpha}_p \,.
\end{multline}
Now $\Omega(X_\zeta,n)$ can be calculated explicitly using the position representation:
\begin{multline}\label{Omega_approx}
  \Omega(X_\zeta,n)
    = \intinfty{}_p\bracket{X,\varphi}{X}_pe^{i\Gamma_S\alpha^2(X-\alpha\sqrt{2})^2}
        {}_p\bracket{X}{\alpha}{}_pdX \\
    = \frac{1}{\sqrt{\pi^{1/2}(\kappa+i\cot\varphi)|\sin\varphi|}}
        \exp\biggl[\frac{1}{\kappa+i\cot\varphi} \hfill \\ \times\biggl(
            -\frac{1+i\kappa\cot\varphi}{2}X_\varphi^2
            + \frac{\sqrt{2}i\alpha\kappa}{\sin\varphi}X_\varphi
            - i\alpha^2\kappa\cot\varphi
          \biggr)\biggr] ,
\end{multline}
where
\begin{equation}
  \kappa = 1 - 2i\Gamma_S\alpha^2 \,,
\end{equation}
see \cite{Schleich2001}, Sec.\,4.4.2.


\end{document}